\documentclass[10pt,conference]{IEEEtran}

\IEEEoverridecommandlockouts

\usepackage[T1]{fontenc}
\usepackage[utf8]{inputenc} 
\usepackage{cite}
\usepackage{amsmath,amssymb,amsfonts}
\usepackage{graphicx}
\usepackage{textcomp}
\usepackage{tabularx}
\usepackage{booktabs}
\usepackage{paralist}
\usepackage{xspace}
\usepackage{tcolorbox}
\usepackage{xcolor}
\definecolor{myblue}{RGB}{235,242,248}
\usepackage{fontawesome5}
\usepackage{balance}
\usepackage{todonotes} 
\usepackage[hidelinks,]{hyperref}

\ifCLASSOPTIONcompsoc
    \usepackage[caption=false,font=normalsize,labelfont=sf,textfont=sf]{subfig}
\else
    \usepackage[caption=false,font=footnotesize]{subfig}
\fi
\usepackage[english]{babel}
\def\BibTeX{{\rm B\kern-.05em{\sc i\kern-.025em b}\kern-.08em
    T\kern-.1667em\lower.7ex\hbox{E}\kern-.125emX}}
    
\begin{document}

\title{%
An Empirical Study of Maintainer-Identified AI Slop Pull Requests in Free and Open Source Software Projects%
\thanks{This research was funded by the CyberExcellence by DigitalWallonia project (No. 2110186), funded by the Public Service of Wallonia (SPW Recherche).%
}%
}

\author{
    \IEEEauthorblockN{Hugo Raskin}
    \IEEEauthorblockA{\textit{University of Namur}\\
    Namur, Belgium\\
    hugo.raskin@student.unamur.be}
    \and
    \IEEEauthorblockN{Xavier Devroey}
    \IEEEauthorblockA{\textit{NADI},
    \textit{University of Namur}\\
    Namur, Belgium \\
    xavier.devroey@unamur.be}%
    \and
    \IEEEauthorblockN{Beno\^it Vanderose}
    \IEEEauthorblockA{\textit{NADI},
    \textit{University of Namur}\\
    Namur, Belgium \\
    benoit.vanderose@unamur.be}%
}

\maketitle

\begin{abstract}
The FOSS ecosystem has always relied on outside contributions, notably through issues or Pull Requests (PRs), to sustain itself.
With the advance of GenAI, the barrier to contributing to FOSS projects has never been lower, leading to the emergence of AI slop PRs: low-quality, low-value generated PRs.
Past studies have examined AI slop PRs through theoretical, platform-level, or practitioner-account frames, but not at the project level.
To this end, this study investigates the incidence of maintainer-identified AI slop PRs in FOSS projects by focusing on PRs flagged as slop by maintainers.
Using a mixed-methods approach, we trace the origins of maintainer-identified AI slop PRs, quantify their volume in FOSS projects, describe their characteristics, and highlight maintainers' reasons for flagging or closing them.
Our results show that the volume of maintainer-identified AI slop PRs increased in mid-2025, peaking in May 2026. These PRs are mostly additive, rarely merged, and typically closed within 3 days. The authors of these PRs disclosed the use of AI in 14\% of cases, with 1\% disclosing after a maintainer's request. 
Most PRs are closed or marked as slop without any reason provided by the maintainer. Based on our open coding, reasons range from concrete elements, like build failures, to ideological or intuitive (i.e., "looks AI to me").
This exploratory study captures a phenomenon in its early stage and highlights GenAI's potential impact on FOSS, emphasizing the pressure it puts on maintainers and the need to shift current open-source development practices.
\end{abstract}


\section{Introduction}
\label{sec:introduction}


Recent advances in generative AI (GenAI) are changing how software is developed, shifting from human-written code to AI-generated code. This shift has several impacts on the Free and Open Source Software (FOSS) community, including an increase in the amount of code produced. 
In response, some communities have limited GenAI use. For example, Codeberg has recently changed its terms of service to ban Large Language Model (LLM)-generated contributions \cite{Codeberg2026ProtectingFLOSS}, and GitHub has announced several new functionalities to limit external contributions \cite{MoraesWolfPullRequestLimits}.

For decades, FOSS has relied on collaborative approaches, enabled by platforms like GitHub, to foster community-based project development where external contributors can propose new features and fix bugs \cite{zhouWhoWillStay2014, tsayLetsTalkIt2014}. In particular, GitHub's Pull Requests (PRs) allow external contributors (\textit{authors}) to propose changes that internal contributors with privileged access (\textit{maintainers}) can review and assess before acceptance (\textit{merging}) into or rejection from the codebase \cite{gousiosExploratoryStudyPullbased2014}. This model has worked for many years, making FOSS a pillar of modern software systems \cite{shen_are_2025}. However, GenAI has lowered the barrier to contributing to FOSS projects by allowing authors with a superficial knowledge of the project to propose changes quickly. For many repositories, this has caused a dramatic increase in the number of PRs to review, leaving the maintainers overwhelmed and forced to drastically adapt their review process. 


In their work, Baltes et al.~\cite{baltesEndlessStreamAI2026} qualitatively analyzed 1,154 posts across 15 discussion threads to understand how practitioners name, frame, and respond to AI slop. Their analysis produced a codebook of 15 codes organized into three thematic clusters: Review Friction, Quality Degradation, and Forces and Consequences. Their results also highlight concerns among developers and propose mitigation strategies for developers, team leads, and educators.
%
In a later work, the same authors frame AI slop in the FOSS ecosystem as a tragedy of the commons~\cite{baltesAISlopSoftware2026}. This theory explains why an individual might overuse a shared resource for personal benefit, leading to its depletion. In this case, the individual's productivity gain comes from AI-generated content in the FOSS ecosystem. Additionally, they propose solutions to the proliferation of AI slop for FOSS decision-makers.

Another study argues that AI slop in the FOSS ecosystem creates a DDoS effect, overwhelming the FOSS community with increasing pressure from generative AI tools~\cite{afrozAISlopDDoSing2026}. Using a mixed-methods approach, they derive hypotheses from public accounts and then evaluate them using Bayesian Structural Time Series analysis across 294 repositories. Their results show an increase in PR volume in 2025, a decrease in merge rates, and a decrease in merge rates for one-time contributors relative to other contributors. Additionally, they present 11 remediation strategies from practitioner interviews, which they then validate with a practitioner survey.

Building on this initial work on the emergence of AI slop in the FOSS ecosystem, our study aims to quantify its incidence more precisely at the repository level. Specifically, we investigate how FOSS projects handle PRs identified as ``AI slop'' and the reasons maintainers give for flagging or closing them. Through this investigation, we seek to understand how maintainer-identified AI slop manifests.


Documenting this phenomenon matters, as AI-related development practices evolve rapidly. While authors and maintainers must continually adapt, researchers face the challenge of capturing evidence before it becomes unavailable. For example, we found that seven PRs we mined from GitHub were no longer available when we conducted our manual analysis. To capture how maintainers identify and respond to this phenomenon,  we focus on PRs explicitly flagged as AI slop by maintainers. We conducted an exploratory study combining GitHub repository mining with manual coding of PRs to answer the following research questions:
\textbf{(RQ$_1$) How do maintainer-identified AI slop PRs impact FOSS projects?}
To answer that question, we focus on the incidence of maintainer-identified AI slop PRs in FOSS repositories over time, and the structural characteristics of such PRs.
\textbf{(RQ$_2$) How do FOSS projects handle maintainer-identified AI slop PRs?}
To answer this question, we annotate the collected PRs for AI disclosure, linked issues, flagging, and closure, and record the reasons maintainers or bots provide for flagging or closing them. We analyze the annotations using descriptive statistics.
\textbf{(RQ$_3$) What reasons do FOSS project maintainers or bots give for flagging or closing PRs as AI slop?}
For this third question, we perform open coding on the reasons collected in RQ$_2$ and use the resulting codes to derive a taxonomy of the reasons for flagging or closing PRs.

The main contributions of this paper are: 
\begin{inparaenum}[(i)]
    \item a \textit{dataset of maintainer-identified AI slop PRs,} supporting future research on the impact of GenAI on FOSS contributions.
    \item \textit{Records of maintainer-identified AI slop PRs,} enabling further study of its impact on FOSS projects.
    \item A \textit{taxonomy of reasons for flagging or closing slop PRs}, which can drive future research and tool development to assist FOSS maintainers.
    \item A \textit{replication package} containing the dataset, analysis scripts, and codebook \cite{replication}.
\end{inparaenum}

\section{Background and Related Work}
\label{sec:background}

\subsection{Collaboration in FOSS Before GenAI}

With the widespread adoption of distributed version control~\cite{barrCohesiveIsolatedDevelopment2012}, pull-based development has become increasingly popular~\cite{gousiosExploratoryStudyPullbased2014}. In this model, a developer proposes changes to a repository by submitting a PR. In a code review process, one or more maintainers (i.e., ``\textit{a role or organization that performs software maintenance activities}'' \cite{washizaki2024swebok}) then review the proposed changes and decide whether to integrate them~\cite{rigbyConvergentContemporarySoftware2013a}. This aims to prevent defects, improve code readability, and share source-code knowledge~\cite{badampudiModernCodeReviews2023, sadowskiModernCodeReview2018}. Code review via PRs is critical in FOSS communities as external contributions form a central part of development~\cite{zhouWhoWillStay2014, shen_are_2025}.


In FOSS projects, external contributors typically lack direct write access to the repository. Beyond technical correctness, maintainers assess whether contributions align with project conventions, design, and software architecture~\cite{tsayLetsTalkIt2014, alamiPullRequestGovernance2022, rigbyUnderstandingBroadcastBased2011}. PR acceptance also depends on social and organizational factors, including prior community interactions, responsiveness during review, and the contributor's reputation~\cite{tsayInfluenceSocialTechnical2014, gousiosWorkPracticesChallenges2015, calefatoPreliminaryAnalysisEffects2017}.

As pull-based development and its supporting platforms have facilitated collaboration, the growing number of open PRs has placed increasing demands on maintainer review capacity~\cite{gousiosWorkPracticesChallenges2015}. PR decisions, whether acceptance or rejection, can be delayed, which adds to the strain on maintainers, already exposed to stress and burnout risk from unhealthy interactions~\cite{ramanStressBurnoutOpen2020a} and sustained maintenance labor~\cite{linakerSustainingMaintenanceLabor2024}, and forces contributors to wait for review~\cite{yuWaitItDeterminants2015}.

\subsection{Collaboration in FOSS After GenAI}

GenAI has made the tension between maintainer capacity and PR review even more pressing as developers increasingly use them in software development~\cite{houLargeLanguageModels2024, fanLargeLanguageModels2023}. LLMs can generate code changes~\cite{chouchenHowSoftwareDevelopers2024}, PR descriptions~\cite{xiaoGenerativeAIPull2024b}, and even code reviews and suggestions~\cite{cihanAutomatedCodeReview2025a}. These AI-assisted PRs lower the barrier to contribution and let new contributors participate, but they can also slow coordination in collective development efforts~\cite{songImpactGenerativeAI2026a}. Autonomous agents now also generate and open PRs entirely on their own~\cite{liRiseAITeammates2025}.

%

Maintainers have increasingly reported instances of ``AI slop''~\cite{BleepingComputer2026CurlBugBounty, Log4jDiscussion4052, Verschelde2026GodotAISlop}, referring to low-value or low-quality AI-generated content. The term generally describes unwanted content that is produced rapidly and in large quantities. Public discourse first framed AI slop in the context of social media, political communication, and cultural pollution~\cite{madsenWhenAITurns2026, klincewiczSlopagandaInteractionPropaganda2025}. AI slop, and vibe coding more broadly, raise concerns about the sustainability of FOSS~\cite{korenVibeCodingKills2026, baltesEndlessStreamAI2026}.



\section{Methodology}
\label{sec:methodology}

\begin{figure*}[t]
    \centering
    \includegraphics[width=0.9\textwidth]{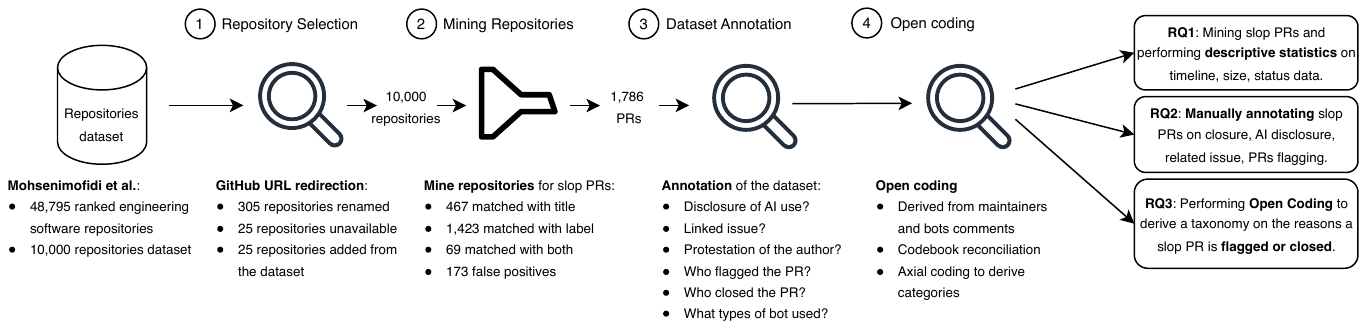}
    \caption{Methodology steps followed to answer the research questions.}%
    \label{fig:method_diagram}%
    \vspace{-5mm}
\end{figure*}

As illustrated in Figure \ref{fig:method_diagram}, we answer the research questions in four steps: we select and mine GitHub repositories to retrieve slop PRs, manually annotate them, and perform open coding based on PR discussions. We provide all collected data, scripts, the annotation procedure, and the open-coding codebook in our replication package~\cite{replication}.

\subsection{Repository Selection}

We focus our mining on the 10,000 software repositories collected by Mohsenimofidi et al.~\cite{mohsenimofidiContextEngineeringAI2026}. They used the SEART GitHub search tool~\cite{dabicSamplingProjectsGithub2021} to collect software engineering projects and applied several criteria, such as number of contributors, creation date, programming language, and compliance with open-source licenses, to focus on actively maintained projects with a long development history. Their analysis yielded 48,795 repositories, which they further ranked by popularity and maturity. Finally, they limited their dataset to the 10,000 most popular and mature repositories on GitHub. Our study is based on their dataset, as its filters allow us to focus on actively maintained and popular software repositories. 
However, although Mohsenimofidi et al.'s dataset was only a few months old, some repositories were already unavailable, either because they were renamed or no longer exist. For renamed repositories, we updated the dataset based on the GitHub redirection from the old URL to the new one. We identified 305 renamed repositories and 25 unavailable repositories. For each unavailable repository, we added the next repository in their initial list until we reached a total of 10,000 repositories to analyze.

\subsection{Mining Repositories}
\label{sec:mining}

\begin{figure}[t]
    \centering
    \includegraphics[width=0.8\columnwidth]{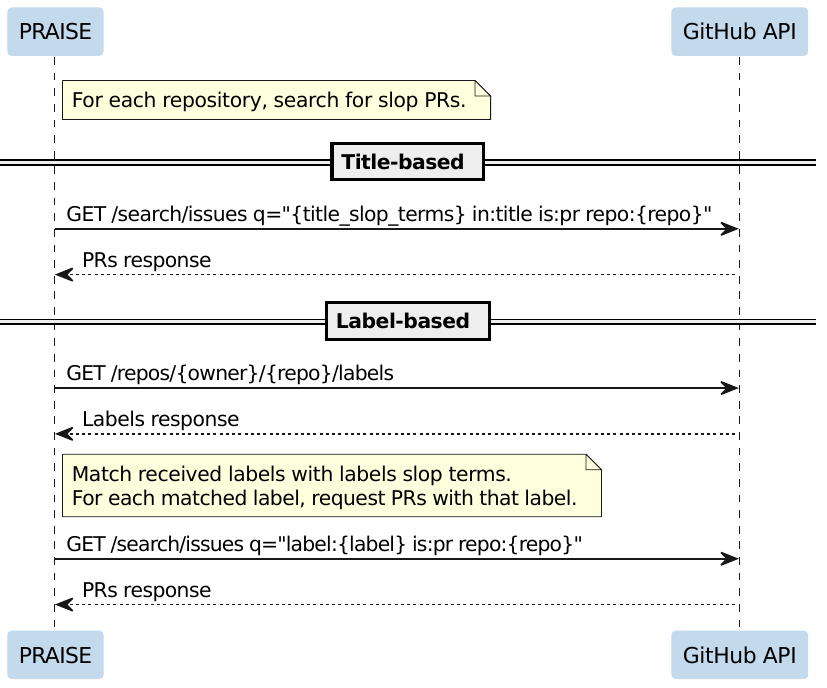}
    \caption{Query to the GitHub API to retrieve slop PRs. The upper part shows the title-based search. The lower part shows the label-based search.}%
    \label{fig:search_method}%
    \vspace{-5mm}
\end{figure}

To answer RQ$_1$, we mined the 10,000 repositories using the \href{https://api.github.com}{GitHub API} to collect PRs flagged as ``AI slop.'' We consider a PR flagged if either of the following holds: 
\begin{inparaenum}[1)]
    \item the PR title mentions slop, or 
    \item the PR carries a label mentioning slop. 
\end{inparaenum}
Figure~\ref{fig:search_method} represents the interactions between our Python script and the GitHub API to retrieve slop PRs based on their title (upper part) or their labels (lower part).

\paragraph{Title-based criterion}
We base our criterion on an initial manual exploration that revealed that maintainers often rename PRs they consider slop, either by replacing the entire title or by adding a prefix. We collect these PRs by querying the GitHub search API using the following keywords, encountered during our exploration, in PR titles: \textit{slop}, \textit{ai junk}, and \textit{ai spam}. We iterate on the repositories by limiting the search query to one repository at a time. 


\paragraph{Label-based criterion}
During the initial exploration, we also observed that maintainers rely on GitHub labels to mark PRs that they consider slop. For each repository, we collect its labels and retain only those that match the following slop terms: \textit{slop}, \textit{ai-pr}, \textit{ai-spam}, \textit{ai spam}, or \textit{rejected ai}. Using the search query, we then query each repository to collect PRs that have at least one matching label.
After collecting the PR information, we made a second call to the \href{https://api.github.com/graphql}{GitHub GraphQL API} to retrieve the number of commits, the comments (reviews, review threads, and comment types), the number of additions and deletions, and author information.

This mining process yielded 1,959 PRs considered slop by maintainers: 467 matched by title, 1,423 matched by label, and 69 matched by both. We quantitatively assessed the construct validity of our mining criteria. To do this, the first author manually reviewed the dataset to identify false positives. For example, in some repositories, the term slop meant something unrelated to low-value generated PRs, or that a bot opened the PR. In total, we identified 173 PRs as false positives, achieving a precision of 91.17\%. We mined from the first PR of each repository through the end of July 2026. After removing the false positives, the final dataset contained 1,786 PRs considered slop by maintainers.

\subsection{Dataset Annotation}

To answer RQ$_2$, we iterated through each PR and used the GitHub web interface to answer following questions:
\begin{inparaenum}[1)]
    \item Does the author of the flagged PR link a related issue?
    \item Does the author of the flagged PR disclose any use of AI? 
    \item Does the author of the flagged PR protest the closing or flagging of the PR? 
    \item Who flagged the PR as slop? 
    \item Who closed the PR?
    \item What type of bot was used to flag or close the PR?    
\end{inparaenum}

If a bot was used, we noted the reason it provided for the flagging or closing, usually described in a PR comment. Lastly, we also noted the reason the maintainer provided for flagging or closing the PR, which we use for open coding (see related section). For the reason for flagging or closing the PR, we only consider comments posted directly in the PR comment feed. To answer the questions, we only consider the state of the PR at the time of the annotation, meaning that if the PR was closed multiple times, we only consider the last person or bot that closed it. To ensure annotation validity, the first author annotated the entire dataset, while the second and third authors each independently annotated half of the dataset. We report the inter-rater agreement using Cohen's $\kappa$ for each question: 1) 0.99; 2) 0.98; 3) 0.90; 4) 1.0; 5) 1.0; and 6) 1.0. All authors discussed any disagreements between annotations.

\subsection{Open Coding}

Finally, to answer RQ$_3$, we open-coded the reasons maintainers and bots gave for flagging or closing the PRs. Following Salda\~na's recommendations \cite{saldana2009coding}, the authors independently coded the PRs based on discussions between the PR author and the maintainers. Comments posted by bots related to the flagging or closing of the PRs were also considered for the coding. A PR can be assigned multiple codes when maintainers gave multiple reasons. 
As in RQ$_2$, the first author coded all PRs, and the second and third authors each coded half. The three codebooks were then reconciled through intensive group discussions among the three authors to define a joint codebook with codes and definitions. We adopt an interpretivist stance and do not report inter-coder agreement, as we treat disagreements as information rather than noise. Lastly, we performed axial coding to group related codes into categories. Each author independently examined the codes and proposed a grouping of related codes. The three proposals were then compared and discussed to reach a consensus on the final categories. Any disagreements were discussed among the authors until an agreement was reached. The category labels were jointly determined through these discussions. 

We documented the whole process, including traceability links and discussions of disagreements, in an Obsidian vault available in our replication package \cite{replication}.

%

\section{Results}
\label{sec:results}


The final dataset contains 1,786 slop PRs mined from GitHub. It is packaged as a MongoDB database in which each PR is represented as a document. The PR document fields fall into three categories: \textit{metadata}, \textit{comment}, and \textit{size}. \textit{Metadata} consists of the PR state (i.e., open or closed), the title and body of the PR, the URL of the PR, whether the PR was merged, and the dates of opening, closing, and last update. They also include information about the PR author, such as login name and GitHub node ID.
\textit{Comment} fields involve information on the discussion of the PR. The raw list of comments is available, alongside comment authors' information. The number of comments posted by bots, by the author of the PR, and by other humans (other than the author). To differentiate comment types, we used data available from the GitHub API response, as GitHub marks comments originating from bots with a boolean field. To distinguish comments from the author and other developers, we simply compared the PR author ID to the comment author ID.
Finally, \textit{size} fields focus on PR-size metrics, such as the number of additions and deletions, the number of commits, and the number of files changed.

\subsection{Impact of Maintainer-identified AI Slop PRs (RQ$_1$)}

To answer RQ$_1$ we analyze the dataset to describe the incidence of slop PRs and their characteristics. 
All analysis scripts are available in the replication package \cite{replication}.


\begin{figure}[t]
    \centering
    \includegraphics[width=0.95\columnwidth]{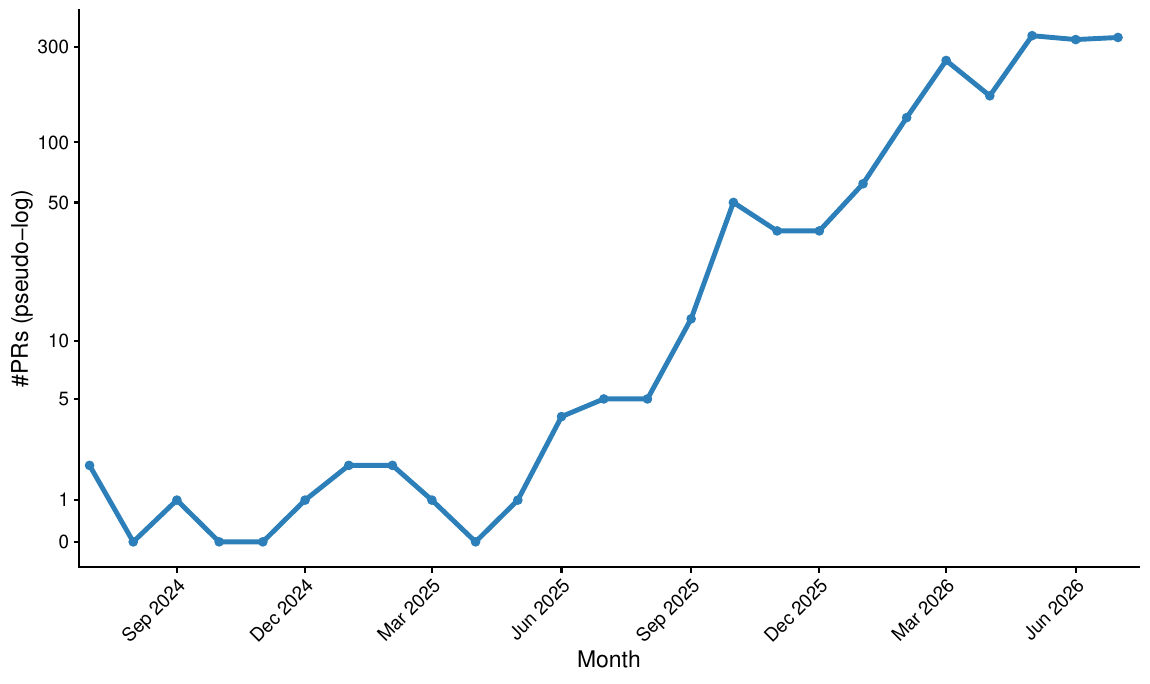}
    \caption{Number of maintainer-identified AI slop PRs (pseudo-log.c scale) opened per month from the first opened PR to the end of July 2026.}%
    \label{fig:pull_requests_evolution}%
    \vspace{-5mm}
\end{figure}

\paragraph{The incidence of slop PRs in FOSS repositories} 
We first present the temporal evolution of maintainer-identified AI slop PRs in FOSS. Figure~\ref{fig:pull_requests_evolution} represents the monthly evolution of the number of opened PRs flagged as slop across all repositories. 
It uses a pseudo-logarithmic scale to better represent variation between months, including months with no PRs flagged. The observation period ranges from July 2024, the month the first PR was flagged, to the end of July 2026.

Between July 2024 and September 2025, the number of PRs opened by month stayed low, ranging from 0 to 5. In September 2025, the number increased from 13 opened PRs marked as slop to 50 the next month. The rest of 2025 was less impacted by slop PRs, with only 36 PRs flagged in November and December 2025. 
In 2026, new slop PRs increased significantly, rising from 62 PRs in January 2026 to 335 PRs in the last month of the observation period (July 2026). The maximum number of open slop PRs was reached in May 2026 with 342 PRs. The last three months of the observation period (May, June, and July 2026) stayed stable in the number of opened slop PRs, with 342, 327, and 335 slop PRs, respectively.



\begin{figure}[t]
    \centering
    \includegraphics[width=\columnwidth]{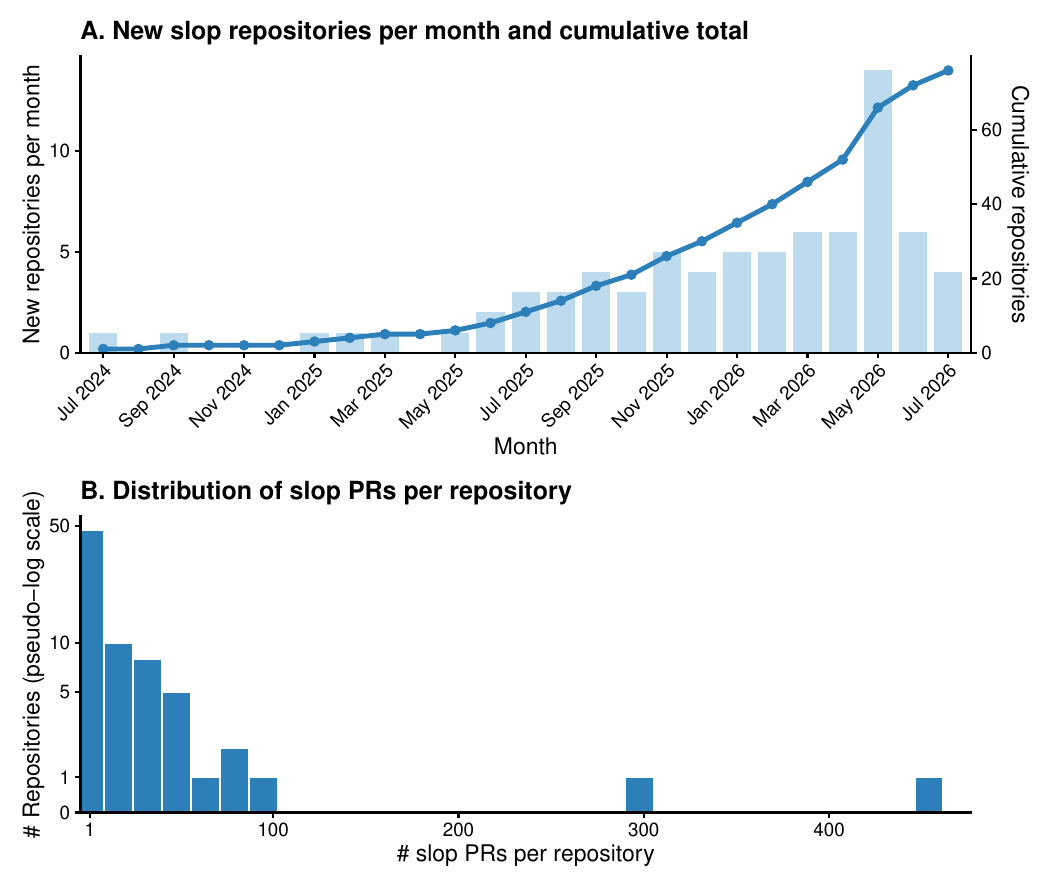}
    \caption{Incidence of slop PRs at the repository level. (A) Monthly number of new repositories affected by slop, based on PR opening date (bars, left axis), and cumulative number of affected repositories (line, right axis). (B) Distribution of the number of slop PRs per repository (pseudo-log scale).}%
    \label{fig:repositories_evolution}%
    \vspace{-5mm}
\end{figure}

To further analyze the phenomenon at the repository level, Figure~\ref{fig:repositories_evolution} shows the incidence of slop PRs by repository. Panel (A) shows a bar chart of the monthly evolution of the number of repositories affected by slop for the first time. The line chart represents the cumulative evolution of new repositories. The date of the opening of the first slop PR in each repository is used to determine when a repository is impacted. We aggregated the opening dates by month to show monthly evolution. Panel (B) shows the number of PRs flagged as slop per repository. The y-axis (number of repositories) uses a pseudo-logarithmic scale.

With the observation period going from July 2024 to July 2026, we observe 76 distinct repositories affected by slop PRs. Between July 2024 and May 2025, the monthly number of new repositories affected by slop PRs stayed low, ranging only from 0 to 1 new repository. However, between July 2025 and April 2026, the monthly number increased, ranging from 2 to 6 new repositories per month. In contrast, May 2026 saw a sharp increase from the previous month, with 14 new repositories affected by slop PRs, making it the month with the highest number of new repositories. June and July 2026 stayed within the range observed before May 2026, with 6 and 4 new repositories, respectively. 

Regarding the number of slop PRs per repository, Figure~\ref{fig:repositories_evolution}. Panel (B) suggests a right-skewed distribution, further confirmed by the statistics (mean $\pm$ standard deviation: $23.50 \pm 62.61$; median: 4.50; min: 1; max: 454). The distribution is concentrated near the lower end of the range, with most repositories having fewer than 5 slop PRs (median: 4.50), while a small number of outlier repositories, reaching up to 454 PRs, create the long right tail and pull the mean well above the median. As the histogram shows, two outlier repositories, \textit{pallets/click} and \textit{huggingface/transformers}, stretch the right tail with 291 (16.29\%) and 454 (25.42\%) slop PRs, respectively. The replication package includes the complete list of 76 repositories and their respective numbers of slop PRs.

\begin{figure}[t]
    \centering
    \includegraphics[width=0.9\columnwidth]{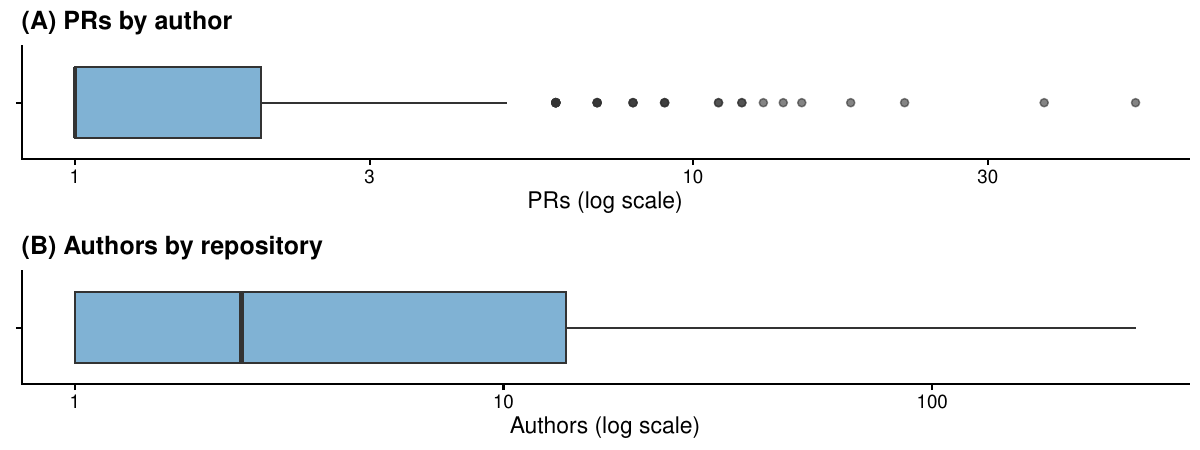}
    \caption{Box plots of slop PR authorship (log scale).
    (A) shows the number of slop PRs opened by the same author across all repositories. 
    (B) shows the number of distinct authors who have opened slop PRs by repository.}%
    \label{fig:prs_by_author}%
    \vspace{-5mm}
\end{figure}

Finally, Figure~\ref{fig:prs_by_author} focuses on the authors of the PRs flagged as slop. Box plot (A) shows the number of PRs, across all repositories, opened by distinct authors, while (B) shows the number of distinct authors who have opened a slop PR by repository. We used GitHub node IDs to identify authors.

Across the dataset, a total of 1,074 distinct authors have opened at least one PR flagged as slop. The statistics show that most authors have opened fewer than two PRs (mean $\pm$ standard deviation: $1.66 \pm 2.48$; median: 1; min: 1; max: 52). With the core of the distribution in a tight window, a group of outlier authors who have opened between 2 and 52 PRs (third quantile to max) stands out. Statistics on the number of distinct authors who have opened a slop PR by repository show that most repositories have slop PRs opened by fewer than three distinct authors (mean $\pm$ standard deviation: $15.88 \pm 45.13$; median: 2.50; min: 1; max: 298). As mentioned, by number of slop PRs per repository, \textit{pallets/click} and \textit{huggingface/transformers} also have the most distinct authors, with 256 and 298, respectively.


\begin{table}[t]
\centering
\scriptsize
\caption{Raw descriptive statistics of slop PRs features. Histograms are on a pseudo-logarithmic scale.}
\vspace{-3mm}
\label{tab:raw_pr_feature}
\resizebox{\ifdim\width>\linewidth\linewidth\else\width\fi}{!}{
\begin{tabular}[t]{p{14mm}p{18mm}rrrrrc}
\toprule
Feature & Description & 5\% & Mean & SD & Med. & 95\% & Histogram\\
\midrule
\textbf{Comment features} &  &  &  &  &  &  & \\
\textit{com\-ments\_to\-tal} 
    & \#Comments (bot and user). & 0 & 2.26 & 3.91 & 1 & 8 & \includegraphics[width=1.5cm]{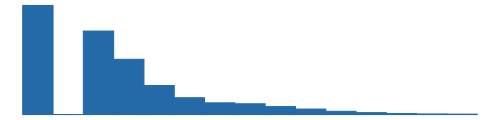}\\
\textit{com\-ments\_fr\-om\_bot} 
    & \#Comments from bot. & 0 & 0.91 & 1.91 & 0 & 4 & \includegraphics[width=1.5cm]{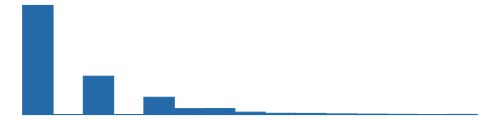}\\
\textit{com\-ments\_not\-\_fr\-om\_author} 
    & \#Comments from users $\neq$ author. & 0 & 1.83 & 3.01 & 1 & 7 & \includegraphics[width=1.5cm]{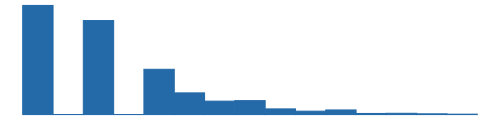}\\
\midrule
\textbf{Size features} &  &  &  &  &  &  & \\
\textit{commits} 
    & \#Commits. & 1 & 9.45 & 237.58 & 1 & 7 & \includegraphics[width=1.5cm]{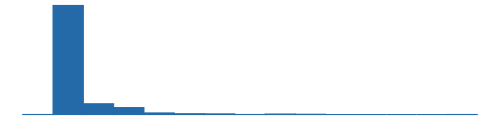}\\
\textit{changed\_files} 
    & \#Changed files. & 1 & 15 & 146.89 & 2 & 22 & \includegraphics[width=1.5cm]{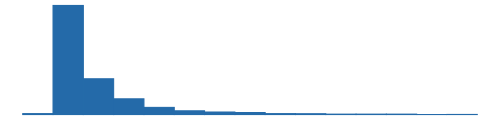}\\
\textit{additions} 
    & \#Lines added. & 1 & 1,413.77 & 15,526.66 & 33 & 1,258.5 & \includegraphics[width=1.5cm]{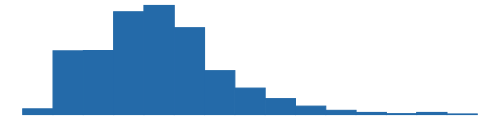}\\
\textit{deletions} 
    & \#Lines deleted. & 0 & 368.92 & 8,469.57 & 2 & 143.5 & \includegraphics[width=1.5cm]{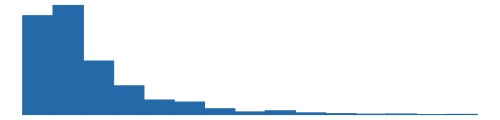}\\
\textit{src\_churn} 
    & \#Lines added + deleted. & 2 & 1,782.69 & 18,223.63 & 37 & 1,575.75 & \includegraphics[width=1.5cm]{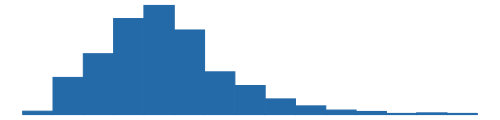}\\
\bottomrule
\end{tabular}}
\vspace{-5mm}
\end{table}

\paragraph{Characteristics of slop PRs}
Table~\ref{tab:raw_pr_feature} presents the overall statistics of the  \textit{comment features} and \textit{size features} for the collected slop PRs. For each feature, we present the 5\% percentile, mean, standard deviation (SD), median, 95\% percentile, and a histogram on a pseudo-logarithmic scale.

The \textit{comments\_total} feature is the total number of comments in the PR (including discussion and review comments). Slop PRs contain, on average, less than 3 comments (mean: 2.26; median: 1). For the \textit{comments\_from\_bot} feature, the results show that PRs contain, on average, less than 1 comment originating from a bot (mean: 0.91; median: 0). \textit{Comments\_not\_from\_author} counts all comments made by users other than the author of the PR. The results indicate that slop PRs contain less than 2 comments from human authors (mean: 1.83; median: 1). However, for each comment feature, the histograms and 95\% percentile indicate a right-skewed distribution, with outliers stretching the tail. The maximum value of each feature, are respectively 47, 31, and 32 comments.


The results for the size features describe highly asymmetric distributions: for each metric, the median is substantially lower than the average, and the average even exceeds the 95\% percentile for the \textit{commits}, \textit{additions}, \textit{deletions}, and \textit{src\_churn} features. This large variation, along with standard deviations far exceeding the means, indicates that a small subset of extreme outliers pulls the average upward. The histograms further confirm this, showing heavily skewed distributions with long tails. The core of the distribution (between the 5\% and 95\% percentiles) is comparatively tight: the majority of PRs contain between 1 and 7 commits, change between 1 and 22 files, and have a total source churn (additions + deletions) of between 2 and 1575 lines.

We further analyze the \textit{additions} and \textit{deletions} features. We define the addition ratio of a PR as $\text{additions} / (\text{additions} + \text{deletions})$. 94.1\% of PRs add more lines than they delete (nonparametric Wilcoxon signed-rank test, $p-value < 2\times10^{-16}$), with a median addition ratio of 0.934, confirming that PRs in this dataset are predominantly additive rather than balanced or deletion-heavy. This pattern remains robust after excluding the top 1\% largest PRs by \textit{src\_churn} size.

Finally, we compare the data distributions of the full dataset and the dataset without PRs from \textit{huggingface/transformers} and \textit{pallets/click}, as they account for 25.42\% and 16.29\% (resp.) using a nonparametric Mann-Whitney U test. All p-values are lower than $0.01$, except for the number of comments from bots (\textit{comments\_from\_bot}, $p-value = 0.5329$) and the number of commits (\textit{commits}, $p-value = 0.5238$). In both cases, the mean increases: from 0.91 (SD=1.91) to 1.06 (SD=2.08) for \textit{comments\_from\_bot} and from 9.45 (SD=237.58) to 14.82 (SD=311.13) for \textit{commits}. However, the median values remain unchanged, denoting potential extreme values bringing the mean values up.

\begin{table}[t]
\centering
\scriptsize
\caption{Slop PRs state distribution and closure time.}
\vspace{-3mm}
\label{tab:pr_state}
\begin{tabular}[t]{lrrrr}
\toprule
State & \#PRs & \% & Mean time & Median time\\
\midrule
Closed (unmerged) & 1748 & 97.9\% & 3.25 days & 0.27 days \\
Merged & 11 & 0.6\% & 13.70  days & 3.33 days \\
Open & 27 & 1.5\% & / & /\\
\bottomrule
\end{tabular}
\vspace{-5mm}
\end{table}

Regarding PR state and closure time, Table~\ref{tab:pr_state} describes the distribution of PR states, mean closure time, and median closure time in days. Most PRs are closed (unmerged): less than 1\% are merged, and 1.5\% were still open at the time of mining. Unmerged PRs close in about 3 days on average, with a median of 0.27 days (about 6 hours). In contrast, merged PRs close in about 13 days on average, with a median of about 3 days. When excluding \textit{huggingface/transformers} and \textit{pallets/click} (outlier repositories), the number of unmerged PRs decreases from 1,748 (97.9\%) to 1,005 (96.5\%). All PRs in those repositories were closed without merging, with only two still open. However, the mean closure time slightly increases from 3.25 days, with a median of ~6.5 hours, to 4.07 days, still with a median of ~6.5 hours. This denotes that the closure time is slightly faster in those two repositories.



\begin{tcolorbox}[
    colback=myblue,
    colframe=myblue,
    boxrule=0pt,
    arc=5pt,
    left=5pt,
    right=5pt,
    top=5pt,
    bottom=5pt
]
\textbf{\faSearch\ RQ$_1$ How do maintainer-identified AI slop PRs impact FOSS projects?}
Slop PRs emerged in July 2024 and remained rare until mid-2025, then rose sharply, peaking at 342 PRs and 14 newly affected repositories in May 2026. Most repositories saw fewer than 5 slop PRs opened by less than 3 distinct authors, though outliers reached 454 PRs and 298 authors. Slop PRs are mostly additive, sparsely discussed, and rarely merged (less than 1\%); 97.8\% are closed, typically in less than a day (median), while merged PRs are integrated into the codebase in about three days (median).
\end{tcolorbox}

\subsection{Maintainer-identified AI Slop PRs Management (RQ$_2$)}
\label{sec:results:rq2}

\begin{table}[t]
\centering
\scriptsize
\caption{Annotation with the number of PRs concerned (\#PRs).}
\vspace{-3mm}
\label{tab:annotation_answer}
\begin{tabular}[t]{lp{46mm}r}
\toprule
Question & Answer & \#PRs\\
\midrule
Linked issue? & \textbf{Yes} (66.6\%); \textbf{No} (33.4\%) & 1779 \\
AI disclosure? & \textbf{Yes} (13.9\%); \textbf{No} (84.5\%); \textbf{After} (1.6\%) & 1779 \\
Author protestation? & \textbf{Yes} (4.1\%); \textbf{No} (95.9\%) & 1779 \\
Flagged by? & \textbf{Author} (0.4\%); \textbf{Bot} (1.7\%); \textbf{Maintainer} (97.5\%); \textbf{Both} (0.4\%) & 1778 \\
Closed by? & \textbf{Author} (4.1\%); \textbf{Bot} (5.9\%); \textbf{Maintainer} (90.0\%) & 1738 \\
Bot type used? & \textbf{coderabbitai} (1.7\%); \textbf{electron-issue-tri\-age} (28.6\%); \textbf{github-actions} (69.7\%) & 119 \\
\bottomrule
\end{tabular}
\vspace{-5mm}
\end{table}

This section presents the results of RQ$_2$, based on the manual annotation of the dataset. 
For each question, Table~\ref{tab:annotation_answer} shows the annotation results, including the total number of PRs associated with each question. After mining, the dataset comprised 1,786 PRs. However, between the mining and the annotation, 7 PRs became unavailable on GitHub. Therefore, we excluded those PRs from the annotation and open coding results. We discussed those deletions in the discussion section.

Our data show that if a majority (66.6\%) of PRs link to a related issue, only 13.9\% of PRs disclose upfront the use of AI. An additional 1.6\% of PRs disclose AI use afterward, either in a comment or by updating the PR description, after a maintainer asks whether AI was used. We used the comment date and description update date to identify such cases. Only a small number (4.1\%) of the authors protest the flagging or closing of their PR, either by questioning the decision, claiming that no AI was used, requesting that the PR be reopened, or even by lashing out at the maintainers.


Regarding closed PRs, we identified 9 PRs where the GitHub interface does not show who closed them. Similarly, for the PR flag question, one PR is flagged by label, but the GitHub interface does not indicate who added the label. These PRs are therefore excluded from our analysis. 
For both questions, maintainers flagged (97.5\%) and closed the PRs (90.0\%) in the majority. The authors closed 4.1\% of PRs themselves, and a bot automatically closed 5.9\%. For PR flagging, the author flagged 0.4\% of PRs themselves, a bot automatically flagged 1.7\%, and both a maintainer and a bot flagged less than 1\%. The last question (bot type used) focuses on the bot types used to close or flag PRs. In the dataset, 119 PRs used such a bot, with \textit{GitHub Actions} used in 69.7\% of the PRs, against 28.6\% for \textit{electron-issue-triage}, and \textit{coderabbitai} accounting for only 1.7\% of the PRs.

Excluding outlier repositories changes the percentages for linked issues, AI disclosure, author protestation, and flagging by less than 2.2\%. However, excluding \textit{huggingface/transformers} and \textit{pallets/click} increases the shares of PRs closed by bots (10.2\%) and authors (6.3\%), with maintainers at 83.5\%, consistent with Table \ref{tab:pr_state}. Bot proportions remain similar: \textit{coderabbitai} (1.9\%), \textit{electron-issue-triage} (32.1\%), and \textit{github-actions} (66.0\%).



\begin{tcolorbox}[
    colback=myblue,
    colframe=myblue,
    boxrule=0pt,
    arc=5pt,
    left=5pt,
    right=5pt,
    top=5pt,
    bottom=5pt
]
\textbf{\faSearch\ RQ$_2$ How do FOSS projects handle maintainer-identified AI slop PRs?}
Slop PRs are linked to an issue in 66.6\% of cases, but only 13.9\% disclose AI use, with an additional 1.6\% doing so after a maintainer's request. In 4.1\% of cases, authors protested the flagging or closure of their PR. Maintainers flag and close most slop PRs, while bots, primarily GitHub Actions, were used to close or flag 119 PRs.
\end{tcolorbox}

\subsection{Slop PRs Flagging and Closing (RQ$_3$)}

\begin{table*}[t]
\centering
\scriptsize
\caption{Codes explaining the reasons PRs are closed or flagged as AI slop by maintainers and bots, grouped by category, with the total number of PRs for all projects and for all projects without huggingface/transformers and pallets/click in parentheses.}
\label{tab:codebook}
\begin{tabularx}{\textwidth}{lXcc}
\toprule
\textbf{\textit{Code}} & \textbf{Description} & \textbf{Maint.} & \textbf{Bots} \\
\midrule
\multicolumn{4}{l}{\textbf{Desk rejection}} \\
\midrule
\textit{AI fingerprint}
    & An AI has signed a commit, a message, or a piece of code.
    & 7 (5)
    & - \\
\textit{Build failure}  
    & The PR could not be built, either by the CI or locally by the maintainer.
    & 42 (40)     
    & - \\
\textit{Insufficient verification}  
    & The PR does not include sufficient elements to show that the changes have been verified (e.g., test files, a proof of test, or decreased test coverage).  
    & 67 (64)
    & 10 (10) \\
\midrule
\multicolumn{4}{l}{\textbf{Bad code}} \\
\midrule
\textit{Could be malicious} 
    & The maintainer suspects that the change would introduce malicious code to the codebase. 
    & 3 (2)
    & - \\
\textit{Incorrect fix} 
    & The code change does not fix the underlying problem. 
    & 64 (56)
    & - \\
\midrule
\multicolumn{4}{l}{\textbf{Already done}} \\
\midrule
\textit{Issue already fixed} 
    & The issue the PR addresses has already been marked as fixed. 
    & 18 (7) 
    & - \\
\textit{Duplicated PR} 
    & The PR is a duplicate of another PR that is still open.
    & 76 (48)
    & - \\
\textit{Superseded by another PR} 
    & The PR has been superseded by another PR: i.e., changes or part of the changes introduced by the PR have been included in the new PR. 
    & 12 (8)
    & - \\
\midrule
\multicolumn{4}{l}{\textbf{Unaware changes}} \\
\midrule
\textit{Issue hijacking} 
    & The issue was opened and assigned to a contributor, but another author hijacked it to address it. 
    & 15 (4)
    & 1 (1) \\
\textit{Speculative change} 
    & The PR tries to fix a fault that never occurs in practice or add a feature that has never been discussed beforehand. 
    & 46 (35)
    & - \\
\textit{Low hanging fruits} 
    & The PR addresses minor issues or nitpicks without added value.
    & 5 (5)
    & - \\
\textit{Pointless change} 
    & The maintainer does not see the added functional or non-functional value of the changes made in the PR.
    & 92 (73)
    & - \\
\midrule
\multicolumn{4}{l}{\textbf{Unreasonable diff}} \\
\midrule
\textit{Inappropriate change size} 
    & The scope of the change is unreasonable, either too large or too small. 
    & 37 (33)
    & - \\
\textit{Low-quality contribution} 
    & The changes are of poor quality and the maintainer points to a specific part of the committed data or artifact, not the PR's title or description. 
    & 36 (32)
    & 4 (4)\\
\textit{Inconsistent changes} 
    & The changes are inconsistent with each other or with the PR's description or title. 
    & 49 (42)
    & - \\
\textit{Preposterous modifications} 
    & The changes do not make sense to the maintainer, and the maintainer points to a specific element in the PR.
    & 122 (99)
    & - \\
\midrule
\multicolumn{4}{l}{\textbf{Repository policy violation}} \\
\midrule
\textit{AI policy violation} 
    & The author did not follow the project's AI policy. 
    & 141 (90) 
    & - \\
\textit{Contribution guidelines violation} 
    & The project's contributor guidelines were not followed, regardless of the AI policy. 
    & 88 (87)
    & 79 (45) \\
\textit{Repository misuse} 
    & The author opened the PR for personal reasons that do not align with open-source principles.
    & 13 (12)
    & - \\
\midrule
\multicolumn{4}{l}{\textbf{Contributor's profile}} \\
\midrule
\textit{Unresponsive author} 
    & The author submitted the PR and did not respond afterward.
    & 20 (16)
    & - \\
\textit{Author's profile} 
    & The PR was closed based on the author's profile. 
    & 82 (68)
    & 12 (-) \\
\midrule
\multicolumn{4}{l}{\textbf{Arbitrary judgment}} \\
\midrule
\textit{Ideological stance} 
    & The maintainer has a strong stance against AI-generated contributions for personal or ideological reasons. 
    & 9 (6)
    & - \\
\textit{Looks AI to me} 
    & The maintainer gave no reason for marking the PR other than that it looked AI-generated. 
    & 117 (79)
    & 54 (54) \\
\midrule
\multicolumn{4}{l}{\textbf{Undisclosed}} \\
\midrule
\textit{Unspecified} 
    & No reason was specified by the maintainer for marking the PR. 
    & 439 (100)
    & - \\
\bottomrule
\end{tabularx}
\vspace{-5mm}
\end{table*}

Table \ref{tab:codebook} presents the final codebook of reasons maintainers and bots close or flag PRs as slop. Each code includes its description and the corresponding number of PRs for maintainers and bots, respectively.
We group codes into categories and sort them from the most concrete to the most abstract criteria used to tag PRs as slop. We detail each code hereafter with examples of PRs. We also report the number of PRs without \textit{huggingface/transformers} and \textit{pallets/click}. The main differences are (i) the drop in the number of \textit{unspecified} PRs, and (ii) the usage of bots based on the \textit{author's profile} in \textit{huggingface/transformers}.

\paragraph{Desk rejection}
This is the most concrete category where maintainers rely on factual elements to mark the PR as slop. 
It includes \textit{AI fingerprints}, where AI has explicitly signed a commit (\href{https://github.com/freqtrade/freqtrade/pull/12819}{freqtrade\#12819}) or left a message in the source code (\href{https://github.com/huggingface/transformers/pull/45076}{transformers\#45076}),
\textit{build failures}, including compilation (\href{https://github.com/electron/electron/pull/51538}{electron\#51538}), continuous integration (\href{https://github.com/apache/airflow/pull/68277}{airflow\#68277}) or quality checks failures (\href{https://github.com/gemrb/gemrb/pull/2249}{gemrb\#2249}), and 
\textit{insufficient verifications} when a ``proof of work'' (\href{https://github.com/bpg/terraform-provider-proxmox/pull/2039}{terraform-provider-proxmox\#2039}) or screenshot (\href{https://github.com/refined-github/refined-github/pull/9561}{refined-github\#9561}) is required to show that the changes have been verified, or if the PR fails to provide evidence of testing (\href{https://github.com/vllm-project/vllm/pull/46672}{vllm\#46672}).

\paragraph{Bad code}
Relying on expertise, PRs are also flagged when they \textit{could be malicious}, e.g., containing committed binary files (\href{https://github.com/autokey/autokey/pull/1167}{autokey\#1167}) or suspicious dependencies (\href{https://github.com/huggingface/transformers/pull/47278}{transformers\#47278}), have \textit{incorrect fixes} due to wrong assumptions (\href{https://github.com/freqtrade/freqtrade/pull/12951}{freqtrade\#12951}) or unchecked code like comments left for future implementation (\href{https://github.com/huggingface/transformers/pull/45076}{transformers\#45076}).

\paragraph{Already done}
Similarly, based on the knowledge of the project, maintainers mark or close PRs when: 
they address \textit{issues already fixed} (\href{https://github.com/OctoPrint/OctoPrint/pull/5169}{OctoPrint\#5169}, \href{https://github.com/apache/airflow/pull/68124}{airflow\#68124}), 
when they are \textit{duplicated PRs} that only contain changes already present in another previously opened PR (\href{https://github.com/huggingface/transformers/pull/44791}{transformers\#44791}, \href{https://github.com/pallets/click/pull/3494}{click\#3494}), 
or when the PR is \textit{superseded by another PR}. This latter case happens when part of the original PR is considered interesting for the project by the maintainer and integrated in another new PR (\href{https://github.com/huggingface/transformers/pull/43330}{transformers\#43330}, \href{https://github.com/zeux/pugixml/pull/693}{pugixml\#693}).

\paragraph{Unaware changes}
This denotes changes submitted without considering the project context. \textit{Issue hijacking} is the most illustrative case: a PR is submitted, addressing an issue that is not assigned to the PR's author (\href{https://github.com/OWASP/Nettacker/pull/1579}{Nettacker\#1579}, \href{https://github.com/huggingface/transformers/pull/46672}{transformers\#46672}) or that is still being debated or discussed on how to address it (\href{https://github.com/huggingface/transformers/pull/45043}{transformers\#45043}, \href{https://github.com/pypa/pip-tools/pull/2432}{pip-tools\#2432}).
A \textit{speculative change} is the opposite: e.g., when a PR tries to fix a problem that cannot happen in practice based on (false positive) static analysis (\href{https://github.com/huggingface/transformers/pull/47389}{transformers\#47389}), code that is actually intended (\href{https://github.com/huggingface/transformers/pull/44652}{transformers\#44652}), or introducing changes, such as new features, that have not been discussed beforehand (\href{https://github.com/vllm-project/vllm/pull/44723}{vllm\#44723}, \href{https://github.com/gnuradio/gnuradio/pull/7956}{gnuradio\#7956}). 
Finally, \textit{low hanging fruits} PRs add small value to the project, e.g., rephrasing documentation or fixing typos (\href{https://github.com/apache/airflow/pull/67758}{airflow\#67758}), while \textit{pointless changes} are for PRs that do not bring any value according to the maintainer, e.g., changes in deprecated code (\href{https://github.com/vllm-project/vllm/pull/41921}{vllm\#41921}), or changing the formatting of the code (\href{https://github.com/microg/GmsCore/pull/3639}{GmsCore\#3639}).

\paragraph{Unreasonable diff}
PRs pointed out for having an \textit{inappropriate change size} are either too big (\href{https://github.com/mesonbuild/meson/pull/15894}{meson\#15894}) to be reviewed by a human or too narrow to bring interesting elements (\href{https://github.com/vllm-project/vllm/pull/40025}{vllm\#40025}). We use this code for PRs where the maintainer explicitly pointed out the size as the reason for flagging or closing the PR.
Maintainer also point PRs bringing \textit{low-quality contributions}, based on code and test quality (\href{https://github.com/searxng/searxng/pull/6209}{searxng\#6209}, \href{https://github.com/jgromes/RadioLib/pull/1735}{RadioLib\#1735}). Unlike cases of \textit{build failures} and \textit{insufficient verifications}, the maintainer uses an authoritative argument and does not point to any additional factual element to tag or close the PR.
Unlike PRs displaying \textit{inconsistent changes} within the code (\href{https://github.com/pypa/pip-tools/pull/2357}{pip-tools\#2357}) or between the changes and the PR's description (\href{https://github.com/pandas-dev/pandas/pull/63093}{pandas\#63093}) for which the maintainer understands the modifications but finds them inconsistent, 
PRs recognized as having \textit{preposterous modifications} do not make sense to the maintainer (\href{https://github.com/huggingface/transformers/pull/46486}{transformers\#46486}: ``\textit{Not sure it makes sense}''). In some cases, it indicates a possible lack of control over AI agents: e.g., PRs for which the GitHub workflow was edited (\href{https://github.com/huggingface/transformers/pull/46455}{transformers\#46455}) or the CI configuration was removed (\href{https://github.com/louislam/uptime-kuma/pull/6137}{uptime-kuma\#6137}) to make the build pass.

\paragraph{Repository policy violation}
We distinguish three policy levels pointed out by maintainers: AI policy, contributor guidelines, and general open-source principles. Depending on the repository, \textit{AI policy violation} can include, e.g., not automatically submitting PRs (\href{https://github.com/pypa/pip-tools/pull/2371}{pip-tools\#2371}), allowing generative AI but requiring a human-written PR description (\href{https://github.com/typescript-eslint/typescript-eslint/pull/12272}{typescript-eslint\#12272}), disclosing the use of generative AI in the PR description (\href{https://github.com/freqtrade/freqtrade/pull/13161}{freqtrade\#13161}), not allowing generative AI for a first-time contribution (\href{https://github.com/huggingface/transformers/pull/47686}{transformers\#47686}), or not allowing generative AI at all (\href{https://github.com/pallets/click/pull/3651}{click\#3651}).
Classical \textit{contribution guidelines violations} include, e.g., not following the PR template (\href{https://github.com/typescript-eslint/typescript-eslint/pull/11426}{typescript-eslint\#11426}), ignoring the project's coding style (\href{https://github.com/mesonbuild/meson/pull/16050}{meson\#16050}), not using proper commit messages (\href{https://github.com/tensorflow/tensorflow/pull/92771}{tensorflow\#92771}), etc.
Finally, \textit{Repository misuses} denote PRs not respecting more general open-source principles, e.g., trying AI tools on a third-party repository (\href{https://github.com/freqtrade/freqtrade/pull/12609}{freqtrade\#12609}), or advertising a tool or service (\href{https://github.com/freqtrade/freqtrade/pull/12987}{freqtrade\#12987}).

\paragraph{Contributor's profile}
\textit{Unresponsive authors} denote cases where the author opened a PR without responding afterward, also referred to as ``\textit{drive-by}'' PRs (\href{https://github.com/huggingface/transformers/pull/46313}{transformers\#46313}, \href{https://github.com/freqtrade/freqtrade/pull/13011}{freqtrade\#13011}). PRs can also be marked and closed based on the \textit{author's profile}. E.g., new accounts opening several PRs in a short amount of time (\href{https://github.com/pypa/pip-tools/pull/2417}{pip-tools\#2417}), which can lead to banning the author (\href{https://github.com/vllm-project/vllm/pull/37389}{vllm\#37389}).

\paragraph{Arbitrary judgment}
Maintainers can also tag and close PRs based on an arbitrary judgment, either by adopting an \textit{ideological stance} and refusing to review AI-generated code (\href{https://github.com/louislam/uptime-kuma/pull/7426}{uptime-kuma\#7426}) or discouraging AI contributions (\href{https://github.com/pandas-dev/pandas/pull/62976}{pandas-dev\#62976}), or by considering that the PR ``\textit{looks AI to me}'' (\href{https://github.com/pallets/flask/pull/6034}{flask\#6034}, \href{https://github.com/pandas-dev/pandas/pull/64820}{pandas\#64820}).

\paragraph{Undisclosed}
Finally, a majority of PRs have been closed and marked by maintainers without any comment or review (\href{https://github.com/huggingface/transformers/pull/45565}{transformers\#45565}, \href{https://github.com/louislam/uptime-kuma/pull/6667}{uptime-kuma\#6667}, \href{https://github.com/pallets/click/pull/3433}{click\#3433}). 
We hypothesize that this latter case is consistent with fatigue, leading maintainers to mark or close PRs without further comment after processing several PRs, but alternative explanations such as an internal policy requiring them not to answer such PRs (``\textit{do not feed the troll}'' policy), or discussions happening via another (potentially private) channel remain plausible.

\paragraph{Bots flagging and closing PRs}
Bots rely on a small set of reasons to flag or close PRs, usually based on the analysis of the changes or PR's metadata. E.g., \textit{github-actions} can automatically close PRs if their description does not include proof of testing, denoting \textit{insufficient verification} (\href{https://github.com/refined-github/refined-github/pull/9562}{refined-github\#9562}), or if the PR's author \textit{hijacked an issue} (\href{https://github.com/OWASP/Nettacker/pull/1579}{Nettacker\#1579}). Similarly, \textit{coderabbitai} is used to detect \textit{low quality contribution} (\href{https://github.com/OWASP/Nettacker/pull/1478}{Nettacker\#1478}: ``\textit{This PR exhibits several hallmarks of a low-value contribution}''). The most frequent \textit{contribution guideline violation} concerns PRs not respecting the PR template or lacking elements in the description, detected using \textit{github-actions} (\href{https://github.com/huggingface/trl/pull/6485}{trl\#6485}) or \textit{coderabbitai} (\href{https://github.com/prowler-cloud/prowler/pull/11762}{prowler\#11762}). A \textit{github-actions} bot is also used to automatically flag PRs based on the \textit{author's profile} (\href{https://github.com/huggingface/transformers/pull/45565}{transformers\#45565}: ``\textit{Common reasons for flagging: New GitHub account; Unusually high number of repository forks in a 24-hour window}''). Finally, PRs are also flagged or closed by bots based on unclear criteria (\href{https://github.com/refined-github/refined-github/pull/9770}{refined-github\#9770}: ``\textit{This looks like an AI-generated PR}''), sometimes asking for additional elements (\href{https://github.com/refined-github/refined-github/pull/9829}{refined-github\#9829}: ``\textit{If you're human and tested it, include a screenshot/video/gif} [...]''), or even reopening a new issue (\href{https://github.com/electron/electron/pull/51335}{electron\#51335}: ``\textit{If you believe your PR was closed in error, we welcome you to resubmit.}'')

\begin{tcolorbox}[
    colback=myblue,
    colframe=myblue,
    boxrule=0pt,
    arc=5pt,
    left=5pt,
    right=5pt,
    top=5pt,
    bottom=5pt
]
\textbf{\faSearch\ RQ$_3$ What reasons do FOSS project maintainers and bots give for flagging or closing PRs as AI slop?}
Most PRs are closed or marked as slop without any reason. Based on our open coding, reasons range from concrete elements, like build failures, to ideological or intuitive ones (``looks AI to me''). A large majority of PRs are also closed for AI policy violations or based on the PR author's profile. 
\end{tcolorbox}
\section{Threats to Validity}


\paragraph{Conclusion validity}
Our analyses are mostly descriptive. Given the sample size (1,786 PRs), statistical significance is uninformative on its own (e.g., $p < 2\times10^{-16}$ for the addition ratio). We therefore privilege proportions (94.1\% of PRs add more lines than they delete). As our features follow heavily skewed distributions, we report medians and percentiles alongside the means.

\paragraph{Construct validity}
``AI slop'' is a recent, unstable notion. We built our mining criteria on an initial exploratory study of the keywords and labels developers actually use, and we quantified the instrument's precision: a manual review of all 1,959 matched PRs identified 173 false positives (91.17\%), though a single author conducted this review. Since we cannot estimate recall as other projects may use different tactics or none at all, our results provide a lower bound on the incidence of slop PRs, grounded in maintainers' judgment. We consider this the main threat to our validity.

\paragraph{Internal validity}
We retrieve only the latest PR state at mining time (e.g., reworked PRs are measured at their last commit). Our identification of slop relies on maintainer judgment, which may conflate AI-generated content with other low-value contributions by inexperienced contributors. Author protestations, observed in only 4.1\% of cases, partially mitigate this. The mid-2025 increase could partly reflect confounders such as ``AI slop'' becoming a community meme, so we interpret trends as correlational rather than causal. For RQ$_3$, 
(i) seven PRs deleted between mining and annotation, and PRs deleted before collection entirely, may cause a survivorship bias toward extreme cases. Quantifying it requires longitudinal mining, left for future work; 
(ii) we code only reasons posted in the PRs and may have missed reasons expressed elsewhere (e.g., potentially private channels), which may inflate the \textit{unspecified} category; 
(iii) two repositories account for $\sim$42\% of the dataset. Our analysis (in \ref{tab:codebook}) shows that \textit{unspecified} PRs are overrepresented and that bots relying on \textit{author's profile} information have been observed in only one repository; 
(iv) maintainers' stated reasons are post-hoc rationalizations that may
favor socially legitimate codes over intuitive ones. Nevertheless, such rationalizations provide a basis for defining stronger criteria for the future.

\paragraph{External validity}
Our incidence estimate (76 of 10,000 repositories) is conditioned on a profile of popular, mature, actively maintained projects \cite{mohsenimofidiContextEngineeringAI2026}, which likely increases both exposure to slop and the likelihood of flagging. 
Our strategy also ties to GitHub's terminology and infrastructure, so incidence figures may reflect the diffusion of this vocabulary as much as the phenomenon itself. We conjecture that our taxonomy of reasons (RQ$_3$) transfers better than our incidence figures (RQ$_1$), but this remains to be verified. 
Finally, our window ends in July 2026 at an apparent peak. We cannot predict whether the volume will stabilize or continue to grow.


\paragraph{Reliability}
Independent annotators double-coded half of the dataset for RQ$_2$ (Cohen's $\kappa$ from 0.90 to 1.00 with disagreements discussed and resolved). The manual review of false positives involved a single author. The list of false positive PRs is documented in the replication package \cite{replication}. 
For RQ$_3$, consistent with an interpretivist stance, we do not report inter-coder agreement and instead treat disagreements as information, relying on a jointly negotiated codebook and fully documented traceability. Since GitHub data evolves continuously, our results are tied to the frozen snapshot in the replication package. Re-running our scripts should be expected to yield different figures.


\paragraph{Ethical Considerations}
We mined only public data from GitHub, without interacting with human participants, complying with the platform's terms of service. Nevertheless, the flagged PRs concern identifiable authors, and our aggregate ``slop'' labeling creates a lasting characterization of these individuals. We therefore limit citations to what illustrates the coding categories and report author behavior at an aggregate level where possible. As slop PRs are often deleted, preserving this snapshot also helps the research community document a transient phenomenon.

\section{Discussion}
\label{sec:discussion}


\subsection{Incidence of maintainer-identified AI slop PRs}

Regarding the incidence within the FOSS ecosystem, our results identify mid-2025 as the tipping point after which slop PRs increased sharply. These results confirm Afroz et al.'s findings ~\cite{afrozAISlopDDoSing2026}. Although our study cannot establish causality, the observed surge temporally coincides with the widespread adoption of agentic coding tools during the course of 2025. Unlike code-completion assistants, agentic coding tools can autonomously identify issues, generate patches, produce PR descriptions, and submit PRs with limited human oversight (``one-click PRs''). This evolution changes the dynamics of FOSS contribution by lowering the barriers to opening a contribution, regardless of its value to maintainers (the occurrences of \textit{Already done} and \textit{Unaware changes} PRs tend to support this hypothesis). The contributors’ motivations for relying on this one-click approach remain unclear, but they might range from a willingness to boost their GitHub profile to simple curiosity about their agents' capabilities, without ruling out a genuine desire to contribute to the project, albeit pursued in an ill-advised manner.

It is also worth noting that although the phenomenon is intense from the affected repositories' point of view, slop PRs are not yet a homogeneous ecosystem-wide phenomenon; only 76 repositories out of 10,000 were identified. These results suggest that factors such as project popularity and issue-tagging practices determine how exposed a repository is.

\subsection{Maintainer-identified AI slop PRs}

Characterizing AI slop PRs requires a non-slop PRs baseline. Gousios et al.~\cite{gousiosExploratoryStudyPullbased2014} offer such information. The study explores the popularity of the pull-based development model with quantitative and qualitative methods. It analyzes 166,884 PRs across 291 projects. This provides a suitable baseline because the original work was published before AI-assisted coding tools emerged.
The comparison reveals both similarities and differences. On the one hand, for the size fields, our slop PRs show similar medians, 1 commit and 2 changed files in both datasets, but higher means: 9.45 commits and 15 changed files in our dataset, compared to 4.47 commits and 12.12 changed files in the baseline. This highlights outliers pulling the distribution, while the core distribution of slop PRs is similar to that of non-slop PRs.  In contrast, the end of slop PRs' life cycle differs dramatically from that of their non-slop counterparts: barely 1\% are merged, compared with 84.73\% for non-slop PRs. The processing time for non-slop PRs is comparable to that of slop PRs, with both typically taking less than one day. However, their outcomes differ substantially. The number of comments is also similar, with an average of 3 comments for slop and non-slop PRs. Based on this observation, we conjecture that the high number of \textit{undisclosed} PRs suggests that maintainers develop uncompromising classification heuristics. This suggests that AI-generated contributions are not as uniformly large and disruptive as one might think but are mostly driven by a few extremely ``bad'' and rapidly discarded PRs.

The results suggest that, at the FOSS level, the phenomenon may be less severe than the disruption perceived by maintainers directly exposed to it. Nevertheless, for maintainers of affected repositories, the feeling of being overwhelmed appears well founded. Our findings also suggest a progressive erosion of maintainers' willingness to engage with such contributions, as reflected in the increasing reliance on rapid rejection, automated filtering, and closures without justification.





\subsection{Maintainers response}


The following summarizes maintainers' responses to AI slop, based on what we observe in the literature and the studied FOSS repositories. First, much work has focused on contribution guidelines and AI policies to regulate AI use~\cite {horaAIPolicyDisclosure2026, chenMakingAIVisible2026, roblesYouCanContribute2026}. These guidelines require, for example, disclosing the use of AI or forbidding its use altogether.
Another response is to automate PR triage. Even before generative AI emerged, past studies focused on creating triage tools to prioritize incoming PRs~\cite{raskinPRECOGPullRequest2026, azeemActionbasedRecommendationPullrequest2020, vanderveenAutomaticallyPrioritizingPull2015}. Additionally, researchers have automated code review, from research~\cite{shiAutomaticCodeReview2019, tufanoCodeReviewAutomation2024} to industrial tools such as CodeRabbit \cite{CodeRabbit2026}.
To reduce the flow of incoming PRs, GitHub now allows maintainers to limit the number of open PRs by user~\cite{LimitOpenPull2026} or even to disable PRs entirely~\cite{NewRepositorySettings2026}. FOSS communities have also begun to adapt, for example, by using trust management systems in which new contributors must be vouched for before contributing~\cite{hashimotoMitchellhVouch2026, GhosttyCONTRIBUTINGmdMain}.

Several categories in our taxonomy are grounded in objective and verifiable criteria, including \textit{build failures}, \textit{insufficient verification}, \textit{issue hijacking}, and \textit{repository policy violations}. These signals can be partially automated, reducing the review burden placed on maintainers. However, a substantial fraction of the reasons identified in our study depend on subjective assessments, such as determining whether a change is \textit{pointless}, \textit{preposterous}, or simply \textit{looks AI to me}. Such evaluations require contextual knowledge of the project and, more broadly, alignment within the community regarding what constitutes a valuable contribution. The fact that the authors themselves flagged 0.4\% of the annotated PRs as AI-generated might suggest that alignment between contributors and maintainers on the issue is possible, provided that authors understand why AI-powered submissions are problematic.

In any case, these responses are the first of their kind, addressing a new systemic phenomenon that poses a real threat to the sustainability of the FOSS ecosystem. Future studies should investigate these strategies and assess their viability for FOSS. However, these investigations might be made more difficult (even impossible) due to the ephemeral nature of the phenomenon: between the mining phase and the annotation phase of our studies, 7 PRs were already deleted by the maintainers or the contributors themselves (hinting that more slop PRs might have been deleted even before the mining phase). Such deletions create survivorship bias and may complicate future longitudinal studies.

\section{Conclusion and Future Work}
\label{sec:conclusion}


In this study, we investigate the emergence of maintainer-identified AI slop PRs in FOSS. We found that the first trace of identified AI slop dates back to July 2024, but the monthly number of PRs does not increase significantly until mid-2025. In total, 76 repositories are affected, with the majority having fewer than 5 PRs. Those PRs are mostly additive, rarely discussed, and rarely merged (in less than 1\%). 97.8\% are closed and typically processed within 3 to 4 days.
We manually annotated the PR dataset and found that 66.6\% link to an issue, while only 13.9\% disclose AI use. An additional 1.6\% disclosed AI use only after the maintainer requested it. In 4.1\% of cases, the author protested the flagging or closing of their PR. Bots were used to automate the flagging or closing in 119 PRs.
Finally, our coding shows that most PRs are closed or flagged for \textit{unspecified} reasons. Other reasons range from verifiable elements (e.g., \textit{Build failure}) to more ideological or intuitive ones (e.g., \textit{Look AI to me}). Most PRs are also closed for \textit{AI policy violation} or based on the \textit{author's profile}.

The main limitation of this study is that our mining relies on an exploratory study of the keywords and labels developers actually used to identify PRs as AI slop, which may not be representative of the broader FOSS ecosystem. Additionally, because the term ``AI slop'' is relatively recent and its usage remains unstable, we treat our results as a lower-bound estimate of the incidence of ``AI slop'' PRs.


Future work should replicate this analysis later to determine whether the prevalence of maintainer-identified AI slop PRs has changed or remained stable over time. Additionally, our open coding revealed several verifiable reasons for flagging or closing PRs as AI slop, which could inform the development of new tools to assist FOSS maintainers. Finally, since this study relies on maintainers' judgments to identify AI slop PRs, future work should investigate whether these judgments can be explained by PR characteristics, such as code quality metrics or the relevance of proposed features to the project.

\smallskip\noindent\textbf{Acknowledgment.}
Generative AI was used as a tool to help create scripts, in brainstorming exercises, in \LaTeX{} document editing, and to improve phrasing.

\balance
\bibliographystyle{IEEEtranS}
\bibliography{biblio}

\end{document}